# ESO User Support and Observation Preparation for VLTI Science Operations

M. Wittkowski*[a], G. Beccari[a], A. Corporaal[b], A. J. Frost[b], O. Hainaut[a], X. Haubois[b], J. Kammerer[a], A. Labdon[b], A. Mérand[a], J. Olofsson[a], A. Pala[a], C. Paladini[b], I. Percheron[a], M. Petr-Gotzens[a], J. Pritchard[a], M. Rejkuba[a], M. Vioque[a], J. Woillez[a]

[a]European Southern Observatory, Karl-Schwarzschild-Str. 2, 85748 Garching bei München, Germany; [b]European Southern Observatory, Alonso de Córdoba 3107, Vitacura, Santiago, Chile

## ABSTRACT

We present recent enhancements to VLTI observing capabilities from a user-oriented perspective, together with new developments in ESO's Observation Handling Tools and end-to-end operations model for VLTI. Key upgrades include GRAVITY+ Coudé guiding with laser guide stars at the UTs, improved limiting magnitudes for GRAVITY and MATISSE, and a new MATISSE narrow-field off-axis mode, all supported by a streamlined, uniform template structure. Phase 2 tools now enable finding-chart generation and the selection of Coudé guide stars and fringe-tracker targets, and will soon offer expanded preparation features and *uv*-coverage visualisation. We summarise usage statistics for snapshot, astrometric, imaging, and time-series observations, outline VLTI scheduling considerations, and present new video tutorials. Starting in 2026, an archive data stream of reduced GRAVITY dispersed visibilities has become available. Complementary community engagement in VLTI operations continues to grow. We close by identifying operational lessons for future interferometric facilities, including the importance of an end-to-end data-flow and user-support model, flexible scheduling, high-level observation-preparation tools, and coordinated community support to keep complex interferometric facilities usable by a broad community.

## 1. INTRODUCTION

The European Southern Observatory's Very Large Telescope Interferometer (VLTI) is operated as a general-user optical/infrared interferometric facility within the integrated operations framework of the VLT. Since achieving first fringes in 2001, the VLTI has evolved from a facility primarily supporting classical snapshot observations into an observatory system enabling increasingly complex programmes, including imaging, astrometry, and time-series monitoring. In previous contributions, we described the initial setup [1] and the evolution of VLTI science operations [2], driven by the second-generation instruments PIONIER [3], GRAVITY [4], and MATISSE [5], as well as the continued development of ESO's Observation Handling Tools and Data Flow System [6,7]. The latter work [2] focused on operational support for flexible baseline-configuration handling, nested scheduling containers, VLTI-specific constraint sets, and dedicated workflows for imaging, astrometric, and time-series observations.

This evolution has been enabled by the scientific and technical capabilities of the VLTI instrument suite. As outlined in the references given above, PIONIER established efficient four-telescope H-band interferometry at the VLTI, while GRAVITY introduced high-sensitivity K-band beam combination, fringe tracking, dual-field operation, and narrow-angle astrometry. MATISSE extended four-beam interferometric observations into the thermal infrared L, M, and N bands.

Together, these instruments have opened a broad range of science cases beyond the traditional domains of stellar astrophysics and star and planet formation. They have supported, among many other results, imaging of evolved-star surfaces [8], studies of massive stars [9] and their circumstellar environments [10], spatially resolved observations of young stellar systems [11], thermal-infrared imaging of active galactic nuclei such as NGC 1068 [12], near-infrared observations of hot dust in the immediate vicinity of AGN [13], measurements of exoplanet masses and spectra [14], and high-precision observations of the Galactic Centre [15].

* mwittkow@eso.org, www.eso.org

The increased sensitivity, imaging capability, and astrometric precision of the second-generation VLTI instruments have also required corresponding changes in science operations. As described in [1,2], VLTI service-mode observing benefits from the same general VLT Data Flow System as single-UT instruments, from proposal preparation to observation execution, quality control, data reduction, and archive ingestion. At the same time, interferometry imposes specific operational requirements, including the selection and relocation of AT baseline configurations, delay-line and shadowing constraints, the use of interferometric calibrators, and the need to efficiently fill the *uv* plane for imaging observations. The introduction of generic AT configuration names, optional alternative configurations, nested scheduling containers, and atmospheric-turbulence constraints has increased service-mode flexibility and improved the match between user science goals and operational execution.

The common thread behind these developments is that increased interferometric capability must be matched by an equally capable user-support and operations model. Higher sensitivity, more complex observing modes, and broader science cases place new demands on observation preparation, scheduling, execution, calibration, data access, and user training. In this contribution, we therefore describe recent VLTI developments from the perspective of the observer and show how ESO's tools, scheduling concepts, archive products, and community interfaces are evolving into a more integrated end-to-end support model. We conclude by drawing lessons for future interferometric facilities.

## 2. VLTI CAPABILITIES FROM A USER PERSPECTIVE

The latest information on the observational capabilities of the VLTI is available through the instrument web pages and user manuals[†]. Here we summarise those capabilities that most directly affect proposal preparation, Phase 2 preparation, and execution constraints.

### The VLTI system

The VLTI operates with two types of telescopes: the fixed 8.2-m Unit Telescopes (UTs), with baselines between 47 m and 130 m, and the movable 1.8-m Auxiliary Telescopes (ATs), offered in different configurations with baselines between 16 m and 202 m.

The performance of the VLTI, including limiting magnitudes and sky coverage, relies heavily on advanced adaptive optics (AO) systems, namely NAOMI and GPAO. When using the ATs, the AO correction is provided by NAOMI [17], the New Adaptive Optics Module for Interferometry. This system delivers a stable and high Strehl ratio, which is crucial for achieving robust fibre coupling and improving the sensitivity and precision of the interferometric data. NAOMI can guide on stars as faint as magnitude 15 in the *G* band, depending on atmospheric conditions, instrument, and observing mode. NAOMI also provides fast chopping capabilities to support instruments such as MATISSE in subtracting thermal background.

When operating the UTs, the AO correction is provided by the GRAVITY+ Adaptive Optics (GPAO) system [16], installed at Paranal in 2024 and 2025[‡]. This system has significantly improved sensitivity, sky coverage, and contrast compared to the previously used MACAO system [18]. GPAO is a high-order, laser-assisted AO system for the UTs, controlling deformable mirrors to enhance fibre coupling and overall instrument sensitivity. It offers multiple modes using either a natural guide star (NGS) or a laser guide star (LGS). The NGS visible mode has a limiting magnitude of Gaia $G_{RP}$ = 12.5. The LGS visible mode is essential when natural guide stars are too faint, enabling high-order correction for targets as faint as Gaia $G_{RP} \approx 17$. However, even when using LGS modes, a natural guide star is still required for low-order corrections. GPAO also provides near-infrared capabilities via the NGS-IR mode (formerly CIAO) and the LGS-IR mode, which analyse light in the *H* and *K* bands, with a limiting magnitude of about 10 in the *K* band.

---

[†] https://www.eso.org/sci/facilities/paranal/instruments/gravity.html ;
https://www.eso.org/sci/facilities/paranal/instruments/matisse.html ;
https://www.eso.org/sci/facilities/paranal/instruments/pionier.html
[‡] https://www.eso.org/public/blog/gravity-leap-vlti/ ; https://www.eso.org/public/news/eso2519

If the science target itself is fainter than the relevant AO guide-star limit, both NAOMI and GPAO offer off-axis Coudé guiding, provided that a suitable guide star can be found within about 50″ of the science target (for the exact limits, see the VLTI user manual available through the instrument pages referenced above).

**The GRAVITY instrument**

GRAVITY is a four-telescope beam combiner operating in the near-infrared K band (1.98–2.40 µm). It consists of two independent interferometers operating simultaneously: the fringe tracker (FT) and the science channel (SC). The fringe tracker analyses the fringe pattern of a target star at high frequency (approximately 1 kHz) and uses this information in real time to correct for atmospheric turbulence and stabilise the fringes of the science channel. All GRAVITY science observations are assisted by fringe-tracker measurements, with the fringe-tracker spectrometer operating at a fixed low spectral resolution of $R \approx 22$.

The science channel records the spectrally dispersed interference fringes across the *K* band and benefits from the stability provided by the fringe tracker, allowing for long integrations. It offers three spectral resolutions: low ($R \approx 22$), medium ($R \approx 500$), and high ($R \approx 4000$). Moreover, when the fringe-tracker reference star is distinct from the science target, GRAVITY provides a direct measurement of the phase of the science target relative to the fringe-tracker target. This observable corresponds to the differential astrometry between the two targets, enabling precise astrometric measurements.

GRAVITY can combine the light from a single source or from two sources simultaneously, offering five observing modes: (1) single-field mode – on-axis, (2) single-field mode – off-axis, (3) dual-field mode – on-axis, (4) dual-field mode – off-axis, and (5) dual-field wide mode. These modes allow different separations between the fringe-tracker and science-channel targets (Figure 1), as detailed in the user manual. In single-field mode, the same target serves as both fringe-tracker (FT) and science-channel (SC) target.

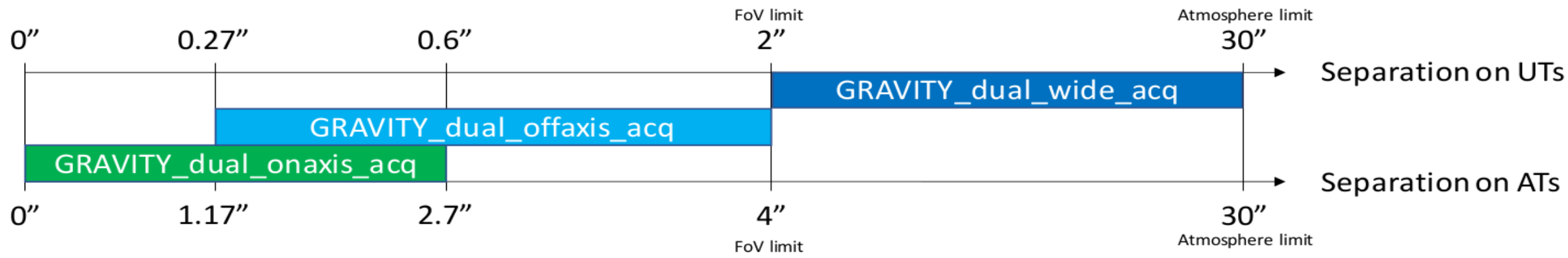

Fig. 1: Possible target separations for the different dual-field acquisition templates: on-axis, off-axis and wide.

In its most sensitive mode, under optimal conditions and with the availability of a fringe-tracker target as faint as $K = 13$ within 2 arcsec of the science target, GRAVITY can reach a limiting *K*-band magnitude of up to 20[§]. The improved capabilities of the VLTI enabled by GRAVITY+ and GPAO have, for example, enabled spatially resolved observations of the broad-line region in a quasar at z = 4 [19]. The GRAVITY+ instrument upgrade is still ongoing with plans to install a new HR-grism and a fiber coupler optics that would improve the astrometric and contrast performance.

**The MATISSE instrument**

MATISSE operates simultaneously in three atmospheric windows: the *L*, *M*, and *N* bands (from 3 to 13 µm). It provides several spectral resolutions: low ($R \approx 30$), medium ($R \approx 500$), high ($R \approx 1000$), and very high ($R \approx 3300$) in the *L* and *M* bands, and low ($R \approx 30$) and high ($R \approx 220$) in the *N* band.

MATISSE can be used in two main configurations:

- **MATISSE standalone**: the classical observing mode in which MATISSE performs its own fringe stabilisation.
- **GRAVITY for MATISSE (GRA4MAT)**: this mode can be used when the correlated *K*-band magnitude of the target is sufficiently bright for fringes to be tracked with GRAVITY. In this configuration, MATISSE observations are carried out with fringes stabilised by the GRAVITY fringe tracker. This allows the use of longer detector integration times (DITs), improving the signal-to-noise ratio in the lower spectral resolution modes and extending the limiting magnitudes to slightly fainter targets, as well as enabling full-band *L* and *M* observations.

[§] https://www.eso.org/sci/facilities/paranal/instruments/gravity/inst.html

A recently introduced specialised GRA4MAT mode is the narrow-field off-axis mode, which enables fringe tracking with GRAVITY while MATISSE observes a nearby offset position. This is particularly useful for observing planets or brown-dwarf companions to stars that can be tracked by GRAVITY. From a preparation perspective, this mode increases the importance of correctly specifying both the MATISSE science position and the GRAVITY fringe-tracking target.

**The PIONIER instrument**

PIONIER, the Precision Integrated Optics Near Infrared ExpeRiment, is a four-telescope *H*-band interferometric combiner at the VLTI, optimised for high-efficiency observations, particularly for interferometric imaging and high-contrast studies. PIONIER offers two main spectral setups:

- **GRISM**: the standard mode, providing limited spectral capability across six channels, with a spectral resolution of $R \approx 30$.
- **FREE**: a single undispersed spectral channel, primarily used for very faint targets to maximise the signal-to-noise ratio.

Further details on recent VLTI system upgrades and future developments are presented elsewhere in this volume [20,21]. Observatory statistics to help the astronomical community become familiar with observatory operations and scientific outcomes is available[**], including information on Phase 1, Phase 2, and publications.

## 3. OBSERVATION PREPARATION

As with all observations at the VLT/VLTI, VLTI observations are prepared using the ESO Phase 2 tool. In this tool, observing blocks (OBs) are defined, specifying the details of the observations. Each VLTI OB includes a target acquisition template and, typically, one science template. The acquisition template contains the information required to point the telescope at a source and to configure the VLTI and instrument for the observation (Figure 2). The common VLTI keywords have been streamlined and aligned across the instruments, simplifying the preparation of VLTI observations. This applies in particular to the Coudé guiding parameters:

- **Adaptive optics mode**: NGS_VIS, LGS_VIS, NGS_IR, LGS_IR, corresponding to the different modes described above.
- **Natural guide star (NGS) source**: SCIENCE or SETUPFILE, specifying whether the science target itself or an off-axis Coudé guide star is used, as discussed above.

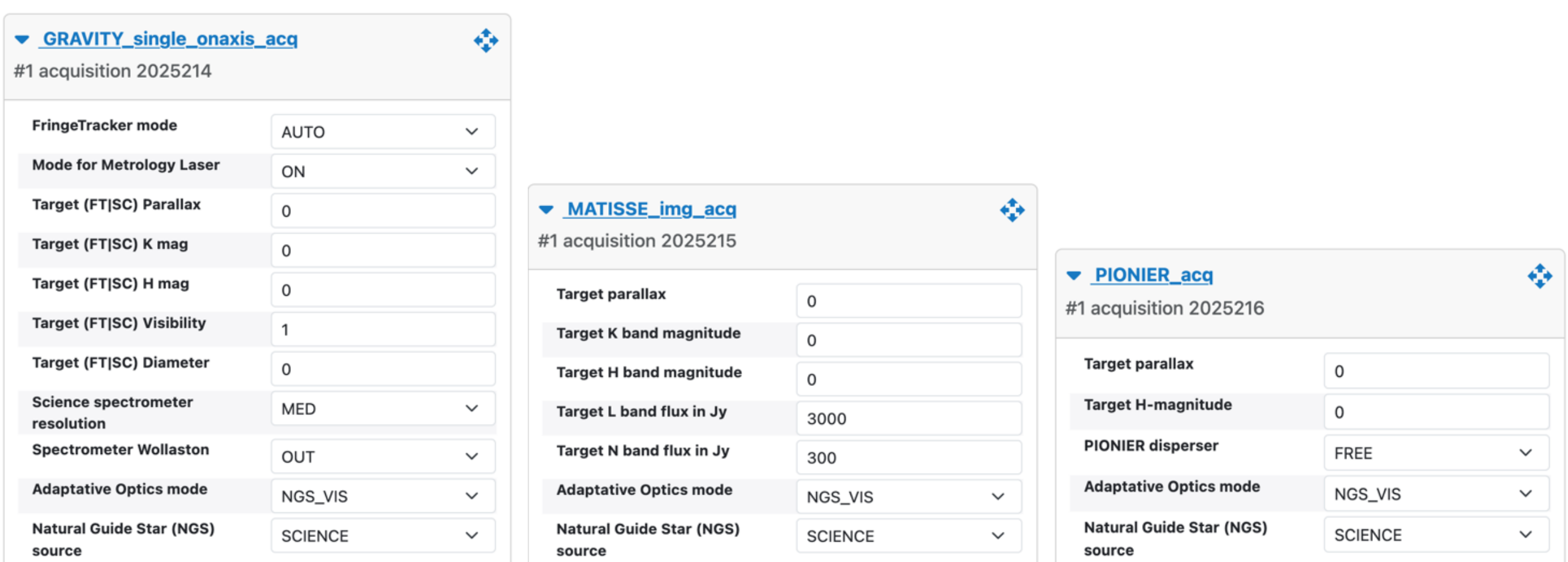

Fig. 2: Acquisition templates for the three VLTI instruments with aligned Coudé parameters

[**] https://www.eso.org/sci/observing/observatory-statistics.html

Phase 2 preparation also includes the definition of the baseline configuration[††], whether using the UTs or one of the AT configurations (Section 6), as well as an observation type tag (snapshot, time-series, astrometry, imaging observation types, see Section 5).

Most interferometric observations must be calibrated by observing a dedicated calibrator star immediately before or after the science target. A calibrator should have a stable, well-known angular diameter or be effectively unresolved. It should also be located close to the science target on the sky and have a comparable magnitude. Tools such as CalVin[‡‡] and JMMC's SearchCal[§§] support the selection of suitable calibrators. Observing blocks are then defined for the selected calibrators using the same instrument mode.

Observation preparation also includes the creation of finding charts to ensure correct target identification during acquisition. Historically, due to limited sensitivity and small fields of view, VLTI targets were predominantly isolated bright sources with little risk of confusion. With the significantly improved sensitivity provided by GRAVITY+ and GPAO, fainter targets in crowded regions have become accessible to the VLTI, making correct target identification more challenging.
The Phase 2 user interface includes a feature that automatically generates finding charts based on 2MASS images. However, for crowded fields, higher-resolution near-IR VISTA/VIRCAM images[***] are often better suited to identify the correct source (Figure 3). In some cases, adaptive optics images, for example obtained through a dedicated pre-imaging run with ERIS, may be required. For such crowded fields, users can customise the background image used to create the Finding Chart within the Phase 2 tool. Possible choices include images available in the ESO Science Archive or mock image generated from the Gaia DR3 catalogue. Automated integration of additional finding charts based on VIRCAM images, where available, is planned.

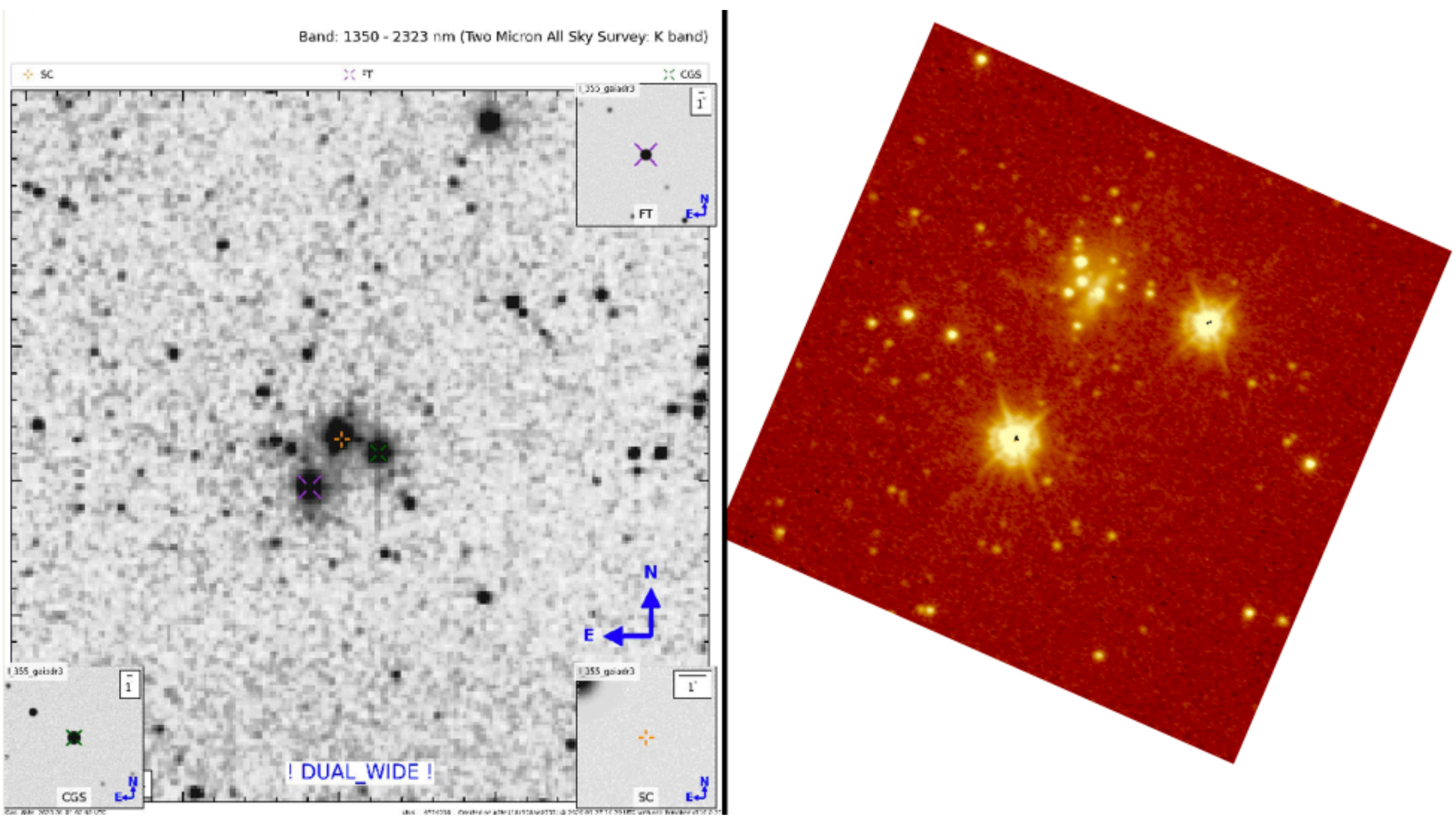

Fig. 3: Example of (left) an automatically generated finding chart from the Phase 2 tool, based on 2MASS imaging, compared with (right) a VISTA/VIRCAM image. In the VISTA/VIRCAM image, similar to what the operator sees at the telescope, the source of interest is resolved into a cluster of sources, making target identification ambiguous.

[††] www.eso.org/sci/facilities/paranal/telescopes/vlti/configuration
[‡‡] https://www.eso.org/observing/etc
[§§] https://www.jmmc.fr/english/tools/proposal-preparation/search-cal/
[***] https://archive.eso.org/scienceportal/home

## 4. THE OBSERVATION PREPARATION (OBSPREP) TOOL FOR VLTI

The increasing complexity of modern astronomical instruments and observing strategies makes it challenging for users to plan and prepare observations, as they often need to provide detailed instrument- and mode-specific information outlined in Sections 2 and 3. To address this, ESO has developed the Observation Preparation tool (ObsPrep), which provides a layer of abstraction over instrument-specific technical details and aims to harmonise the user experience when preparing observing blocks for complex instruments [22].
For VLTI specifically, ObsPrep now supports the selection of Coudé guide stars for the NAOMI and GPAO adaptive optics systems, as well as the selection of a suitable fringe-tracker target (FT) for the GRAVITY dual-field wide mode (Section 2), as illustrated in Figure 4.

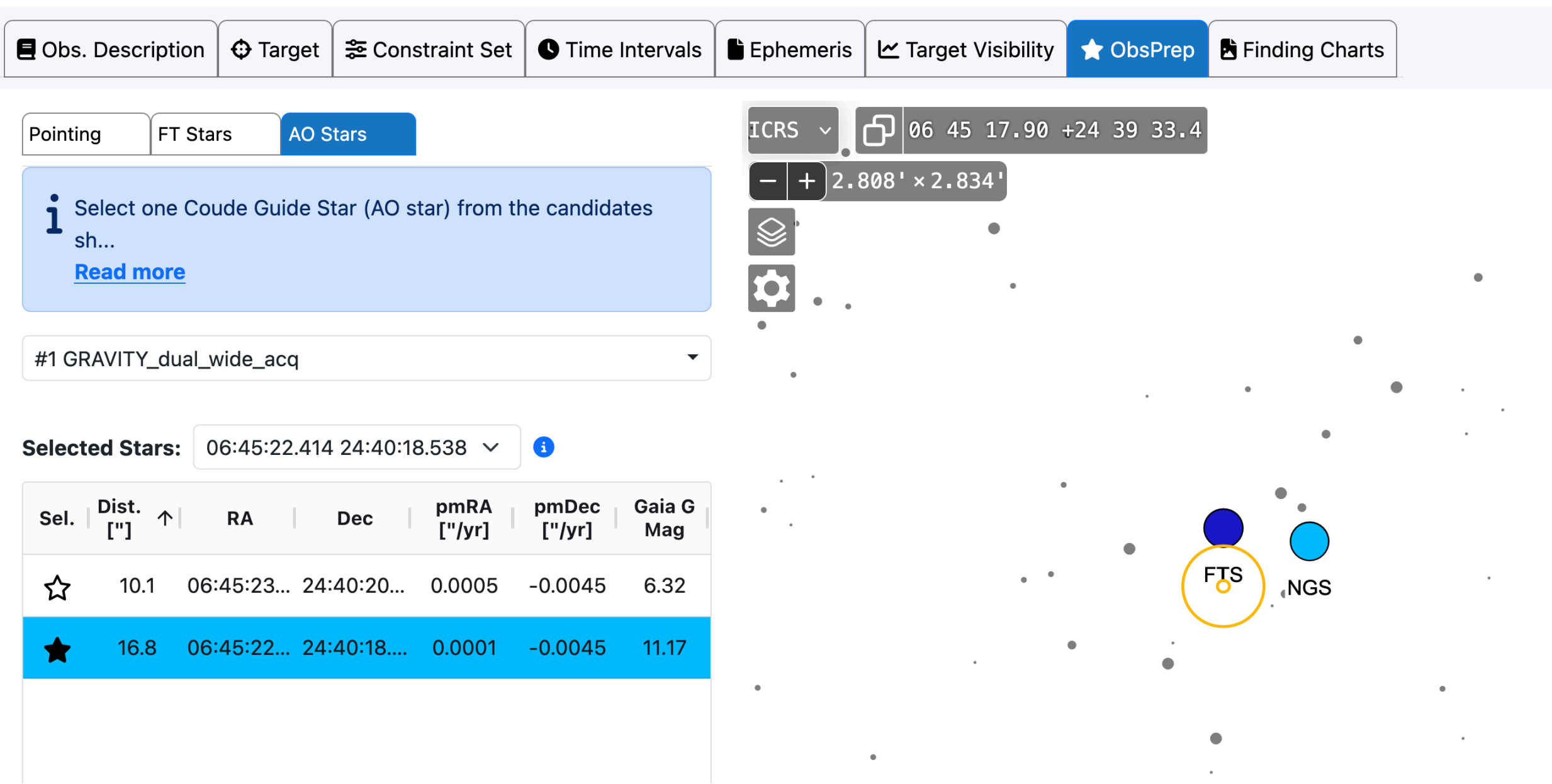

Fig. 4: Example of ObsPrep to select a FT star (FTS) for the GRAVITY dual-field wide mode and a Coudé guide star (NGS)

Planned extensions for VLTI include instrument-mode selection, integration of interferometric calibrator searches, automatic creation of corresponding observing blocks, observability calculations including delay-line restrictions and shadowing, and automatic population of required target information from predefined catalogues.

In this model, most users would no longer need to interact directly with the template structure shown in Figure 2, but instead will use ObsPrep in an intuitive way to prepare observations across VLT, VLTI, and ELT instruments. ObsPrep would then automatically select and populate the appropriate acquisition and observation templates. This approach would significantly simplify observation preparation and make it more accessible to the broader astronomical community.

## 5. USAGE STATISTICS FOR THE DIFFERENT TYPES OF INTERFEROMETRIC OBSERVATIONS

VLTI observations are tagged by one of several observation types, enabling dedicated support for different observing strategies. The distinction between observation types is relevant for operational planning and user support. We therefore summarise recent usage statistics to show how these categories are used in practice and how they translate into different scheduling and support requirements.

- **Imaging** observations use multiple baseline configurations to sample the uv plane for image reconstruction.
- **Snapshot** observations are standalone concatenations without time links or *uv*-plane filling requirements.
- **Time-series** observations involve concatenations repeated over the observing period.
- **Astrometric** observations use GRAVITY dual-field observations to extract relative astrometry.

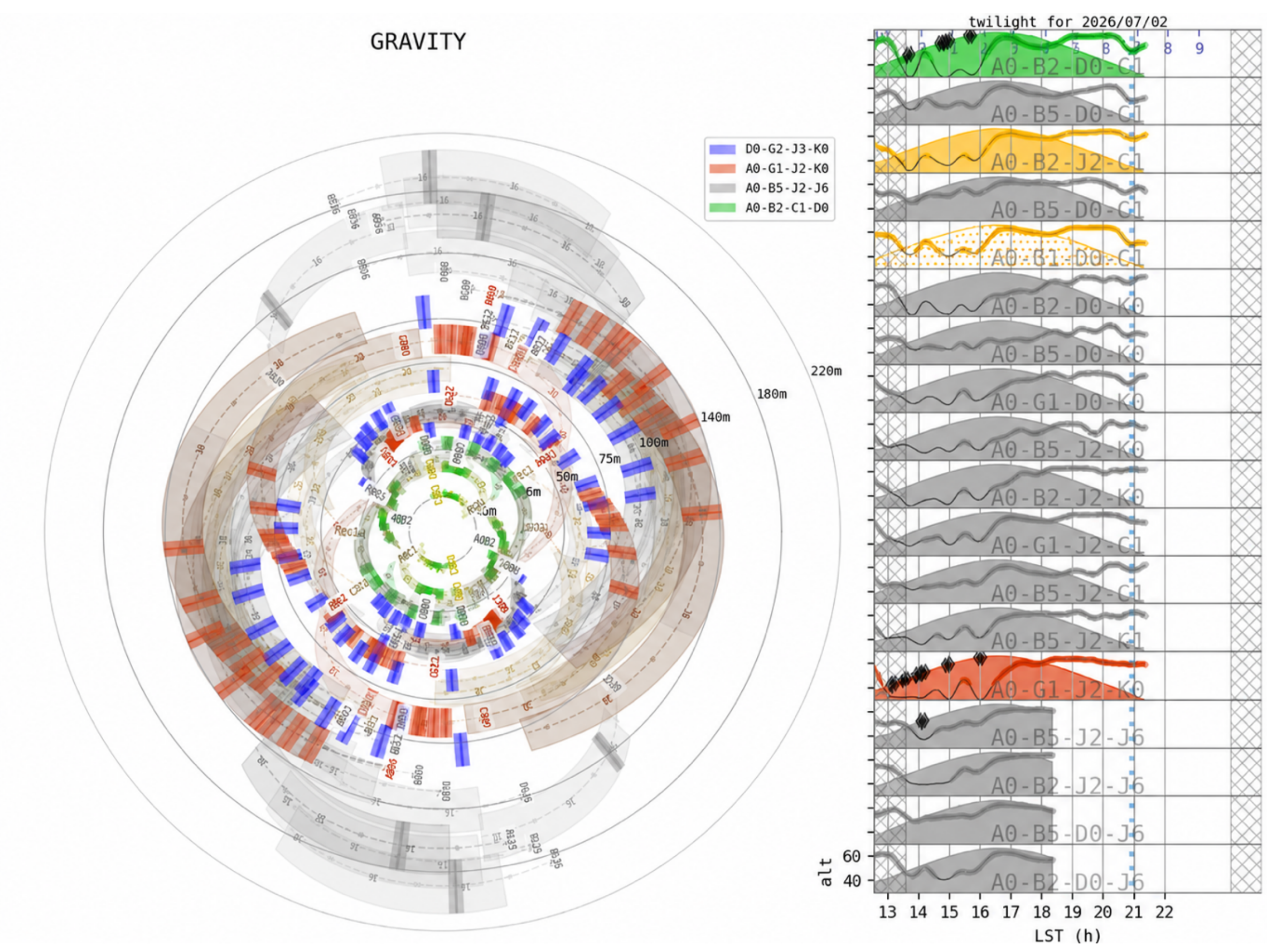

Fig. 5: Example *uv*-coverage achieved for a VLTI imaging observation, combining all four AT configurations (small, medium, large, extended). The left panel shows the uv coverage that can be achieved (shades) and the points already taken (bars). The right panel indicates baseline configurations at local sideral time (LST) that provide additional non-redundant *uv* points.

The different VLTI observation types require different operational strategies and support. Imaging observations require the availability of most or all AT configurations within a limited time window and a scheme to ensure that duplicate *uv* points are avoided during execution, such that the *uv* plane is filled as uniformly as possible. Figure 5 shows an example of the *uv*-coverage tacking graphic used for this purpose. Time-series observations require regular scheduling of specific AT configurations. Due to conflicting requirements between imaging and time-series observations and the need to cycle through different AT baseline configurations (Section 6), time-series observations are typically viable over a few consecutive days, or at intervals of a few months for a given configuration. Astrometric observations require suitable AT baselines (currently excluding the extended configuration) as well as dedicated support for observation preparation and data analysis. Absolute visibility calibration is less critical, and observations of calibrator stars to estimate the interferometric transfer function are often omitted.

Figure 6 shows the evolution of the usage of the different observation types over the last 10 periods (five years), in terms of the fraction of allocated observing runs (left) and allocated time (right), for AT configurations (top) and UTs (bottom). Only A- and B-ranked runs are considered, as C-ranked runs (fillers) are subject to additional selection criteria.
For the ATs, snapshot observations account for the largest fraction of runs in most periods, followed by imaging and time-series programmes. In terms of allocated time, there is a clear decrease for imaging from more than 50% in periods 108–110 to about 20–30% in periods 115–117. Over the same interval, the time allocated to time-series observations increased from less than about 30% to more than 40%. At the time of writing, in period 117, both time-series and imaging observations have a larger time share than snapshot observations. Astrometric observations represent a smaller fraction, with less than 20% of the runs and less than 10% of the allocated time, noting that astrometry is only available with the GRAVITY instrument.
For the UTs, observations are dominated by snapshot programmes, both in terms of fraction of runs and allocated time, followed by astrometry. Time-series and imaging observations account for significantly smaller shares.

Figure 7 shows the completion fractions at the OB level for the different observation types for the ATs. Despite some fluctuations, the completion rates at the OB level are comparable across the different observation types, with no clear indication of systematically different performance between them. Monitoring the usage and completion of the different observation types helps to better understand community needs and to provide tailored support in terms of scheduling and user support.

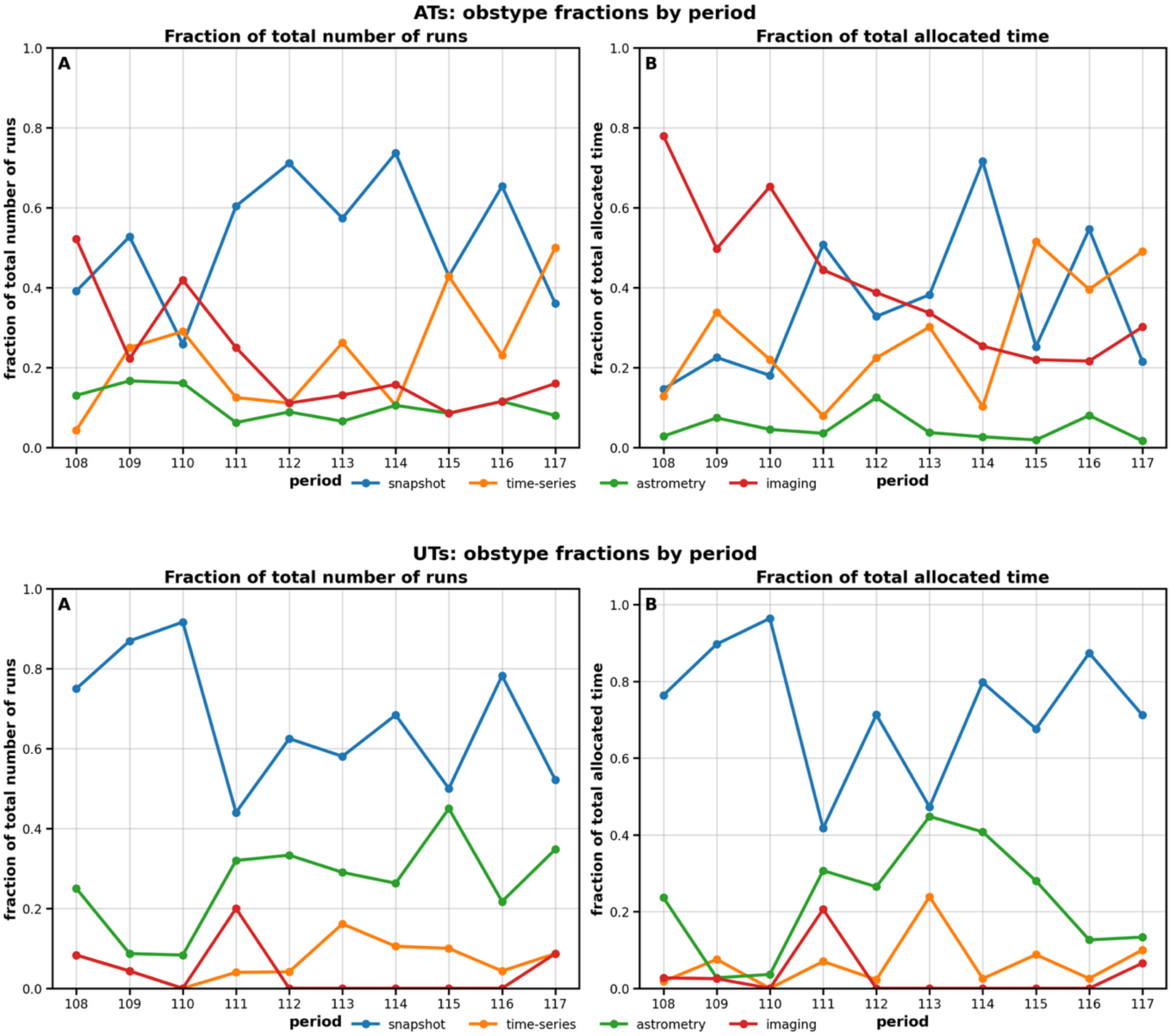

Fig. 6: Usage of different VLTI observation types over the last 10 periods (5 years) in terms of fraction of observing runs (A) and in terms of allocated time (B) for the AT configurations (top) and the UTs (bottom).

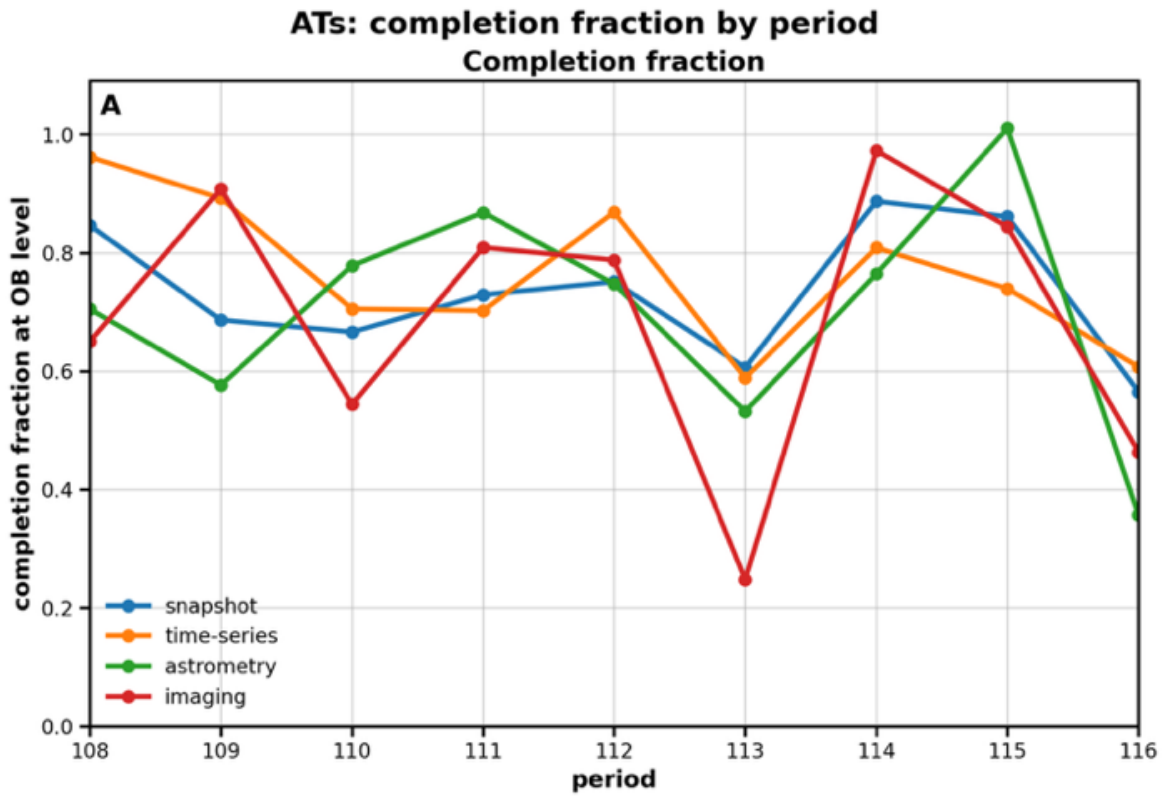

Fig. 7: Completion rates at the OB level for ATs. The completion rate for period 116 will still increase, as some runs were still active at the time of writing.

## Statistics on individual VLTI images

VLTI imaging observations are organised into groups that include all observations required to produce an individual image, enabling monitoring of the number, statistics, and progress of imaging programmes.

Figure 8 shows histograms of the number of *uv* points (left) and the execution time (right) requested for individual images during periods 108 to 117, for A- and B- ranked observing runs. The average number of *uv* points per image is about 14, with an average execution time of about 13.5 hours. A total of 111 images are included, among them several observations with only a small number of *uv* points, which by themselves are insufficient to produce a complete image but may complement earlier observations or data from other interferometers. In the following, we consider only images with at least eight *uv* points, reducing the sample to 81 images, or about eight images per period.

An image is considered completed if at least 80% of the requested *uv* points have been successfully obtained. Figure 9 shows the number of scheduled images (left) and completed images (right) per period. Of the approximately eight images scheduled per period, about five reach the 80% completion criterion. This provides an estimate of the number of images that can realistically be achieved within current scheduling constraints. The completion fraction is lower for time-critical observations (≈50%) compared to those without time constraints (≈70%).

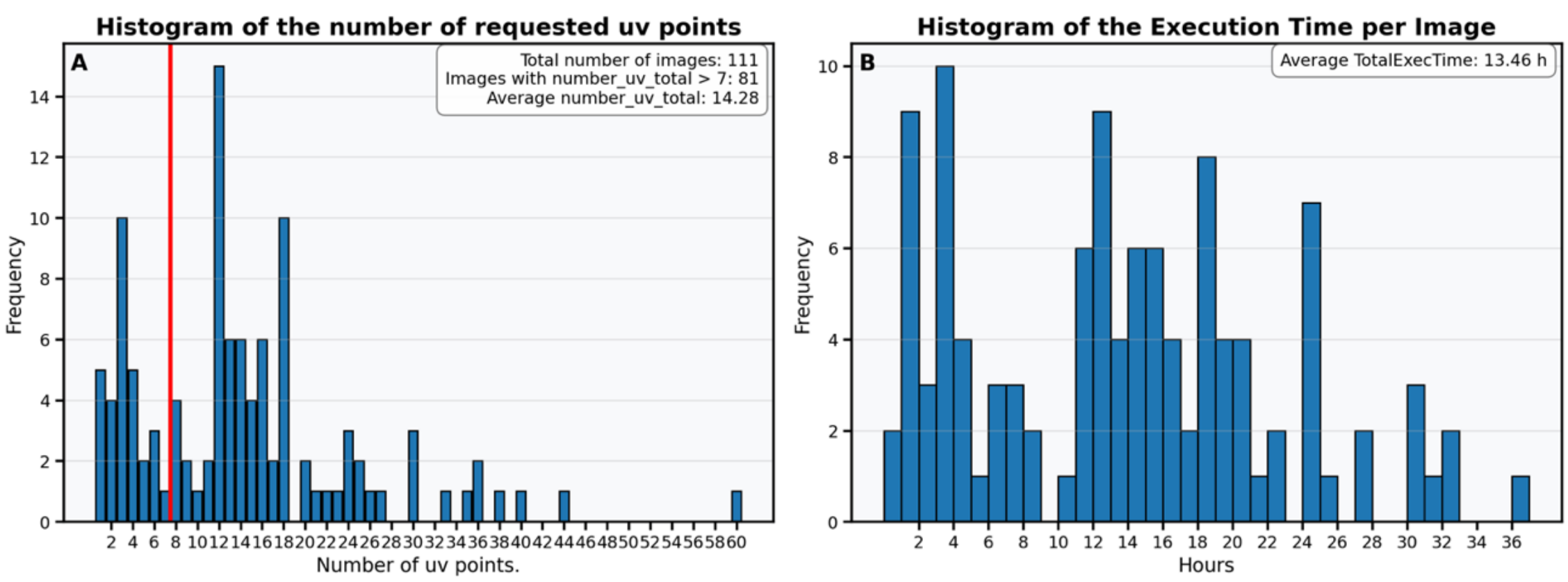

Fig. 8: Histogram of the number of uv points (left) and execution time (right) requested for individual images during periods 108 to 117, for the A- and B-rank class observing runs.

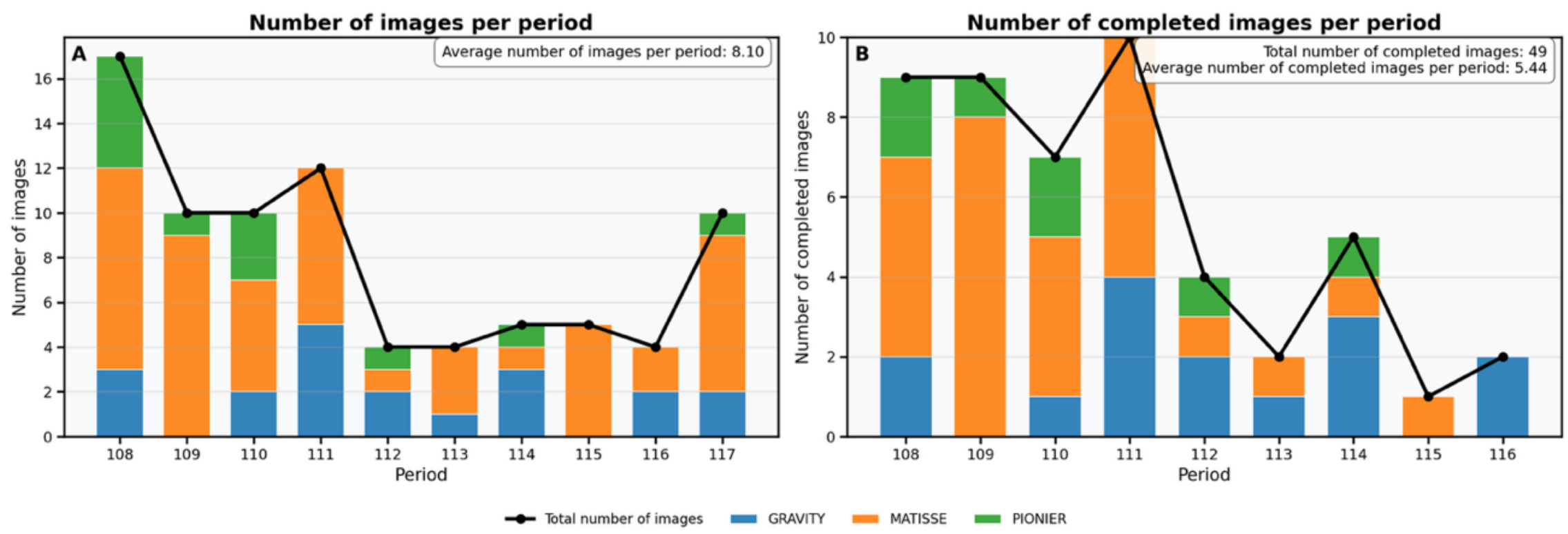

Fig. 9: Number of images scheduled (left) and completed to at least 80% of the requested uv points (right) per period.

**Scientific results based on different VLTI observation types**

In the following, we present a non-exhaustive selection of scientific results based on VLTI data obtained since period 108, focusing on completed imaging, astrometry, time-series, and snapshot observing runs. These results are identified by cross-matching programme IDs with the corresponding observation types and the ESO telescope bibliography†††. These examples demonstrate why the different observation types require tailored support.

Imaging studies [23–28] demonstrate the capability of VLTI to directly resolve complex circumstellar environments at near- and mid-infrared wavelengths:

- Evolved stars and mergers (e.g. V838 Mon, V Hydrae, RT Pav), revealing asymmetric ejecta, dust formation, and large-scale structural complexity.
- Binary interaction and mass transfer (e.g. $\pi^1$ Gru, RT Pav), where imaging resolves circum-companion structures and accretion flows.
- Protoplanetary disks and eruptive systems (e.g. Z CMa), capturing disk morphology changes during outbursts.
- Chemical and spatial stratification in dusty environments (e.g. X TrA).

Astrometric programmes [29–36] exploit VLTI/GRAVITY for companion detection and orbital characterisation:

- Binary and multiple systems: detailed orbital solutions and dynamical masses, for example, for A-type stars and binaries.
- Exoplanet and substellar companions.
- Orbital constraints (e.g. AF Lep b, GQ Lup B).
- Spectro-astrometric characterisation (ExoGRAVITY survey).
- Detection of Gaia companions at extremely small separations.
- Confirmation of embedded protoplanets (e.g. WISPIT 2c).

Time-series interferometry [37–41] captures the temporal evolution of astrophysical systems:

- Dynamical mass measurements of exotic binaries (e.g. Be + stripped star systems).
- Hierarchical systems (δ Circini).
- Microlensing: first direct resolution of lensed images.
- Accretion variability (DX Cha).
- Direct observation of rotating magnetospheres in motion.

††† telbib.eso.org

Snapshot observations [42–47] provide efficient single-epoch spatial and spectral diagnostics across a wide range of targets, enabling the physical characterisation of different classes of objects:

- Star–disk interfaces and inner gaseous regions (e.g. massive YSOs), identifying new tracers such as Na I and He I for accretion and wind-launching zones.
- Mass loss and circumstellar structures in evolved stars (e.g. IRAS 17163−3907), resolving asymmetric outflows and linking mass-loss geometry to stellar evolution.
- Atmospheric structure and wind–atmosphere coupling in red supergiants, constraining extended molecular layers and deviations from hydrostatic models.
- Circumstellar disks and dust environments (e.g. Fomalhaut, post-AGB binaries), probing exozodiacal dust and confirming transition disk configurations.
- Population studies of massive stars, e.g. Wolf–Rayet surveys, providing statistical constraints on binarity, identifying long-period binary deserts, and constraining interactions in massive stellar multiples.

Collectively, these studies span a wide range of astrophysical domains, from star and planet formation to stellar evolution and binary interactions, highlighting the versatility of VLTI observations. The different observing modes provide complementary capabilities: imaging delivers direct spatial information with high visual impact, astrometry enables high-precision measurements and orbital characterisation, time-series observations probe dynamical evolution and transient phenomena, and snapshot observations offer efficient single-epoch spatial and spectral diagnostics, often for large samples of targets.
These examples illustrate that the operational categories used in VLTI support are not merely administrative labels, but correspond to distinct scientific use cases with different requirements on preparation, scheduling, calibration, and data analysis.

## 6. VLTI SCHEDULING ASPECTS

There have been substantial updates to ESO's time allocation and scheduling processes, as well as to the underlying software tools [48]. The system now accommodates both scientific observations and technical time requests within a unified framework. It incorporates new mechanisms for managing commitments across overlapping observing cycles and for accounting for the remaining science-time fraction from previously accepted programmes. New science proposals and updated technical time requests are scheduled within the same optimisation environment. The scheduler also includes the joint optimisation of multiple telescope schedules, including VLTI observations with either the four 8-m UTs or the different AT configurations.

Owing to the need to cover different baseline configurations at various sky positions and observing times in both service and visitor mode, the number of observing nights available for a given baseline configuration at a specific sky position is reduced compared to single-UT instruments. In addition, changing an AT configuration requires physically relocating the telescopes, a process that can take up to two days.

To increase service-mode flexibility for VLTI observations:

(1) AT configurations are requested using generic names (currently "small", "medium", "large", and "extended"), rather than explicit configurations defined by AT positions. This scheme allows more flexible execution of observing blocks, including the use of intermediate configurations that occur during AT relocations. Imaging observations particularly benefit from this approach, as intermediate configurations provide additional *uv*-coverage that would otherwise not be accessible.
(2) Alternative baseline configurations can optionally be specified, if the scientific goals can be achieved with more than one of the available generic configurations (e.g. medium or large). This increases the number of observing blocks that can be executed at any given time.
(3) In addition, the VLTI-AT schedule includes so-called imaging slots, defined as periods of about two weeks of uninterrupted service-mode time, with flexibility in the timing of configuration changes. Imaging slots are primarily intended to support programmes that require multiple baseline configurations within a limited time window, but they are not restricted to imaging observations. They increase overall scheduling flexibility and allow optimisation of the time spent in each AT configuration depending on the progress of the observing programmes.

In this sense, the Phase 2 baseline-configuration choices described above are directly coupled to the operational flexibility provided by the scheduler. The VLT/VLTI telescope schedule, including the planned usage of the different UT and AT baseline configurations, is publicly available[‡‡‡]. An example is shown in Figure 10 for June and July 2026. This information can be used to optimise observation preparation, in particular for time-series and time-critical observations.
These scheduling mechanisms are therefore an essential part of VLTI user support: they translate complex baseline-configuration constraints into executable observing programmes.

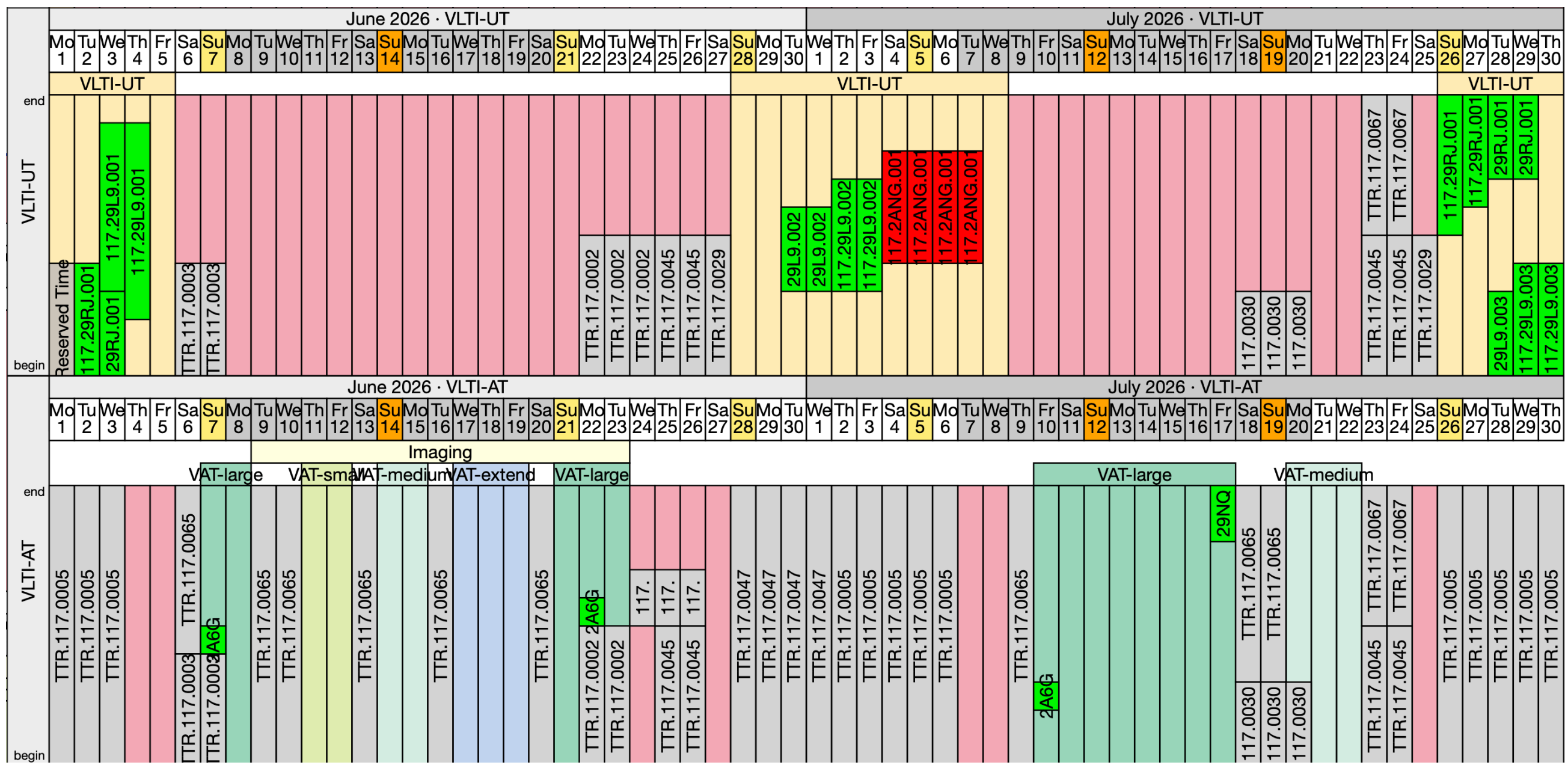

Fig. 10: VLTI UT and AT telescope schedule for June and July 2026, obtained from the public ESO long-term schedule.

## 7. VIDEO TUTORIALS FOR VLTI

User support also relies on documentation and training materials that help users navigate increasingly complex observing modes and workflows. Video tutorials complement the Phase 2 and ObsPrep documentation by lowering the practical entry barrier for VLTI users and making information available in multiple accessible formats and supporting users with different learning preferences.

ESO provides comprehensive information on observing with its telescopes, including telescope time allocation, Phase 1 and Phase 2 observation preparation, and Phase 3 data handling[§§§]. This includes a range of general and instrument-specific web-based tutorials[****], complemented by a helpdesk system for personalised support as well as a Knowledgebase[††††].

The ESO User Support YouTube channel[‡‡‡‡] has now been refreshed with new and updated video tutorials. These guides help users efficiently prepare their Phase 2 material, as well as navigate and query the ESO Science Archive. Complementing the text-based tutorials, they provide an accessible way to explore the observing process at ESO. In particular, the video tutorials target new users unfamiliar with ESO, whereas the text-based tutorials and the Phase 2 documentation are mostly aimed at more advanced users

‡‡‡ https://www.eso.org/LPOschedule
§§§ https://www.eso.org/sci/observing.html
**** https://www.eso.org/sci/observing/phase2/p2intro/p2-tutorials.html
†††† https://support.eso.org/
‡‡‡‡ https://www.youtube.com/@ESOsupport

For VLTI support in particular, four video tutorials have recently been added (Figure 11):
• ESO Phase 2 – An introduction to the VLTI[§§§§]
• ESO Phase 2 – Preparing your PIONIER observations[*****]
• ESO Phase 2 – Preparing your MATISSE observations[†††††]
• ESO Phase 2 – Preparing your GRAVITY observations[‡‡‡‡‡]

Overall, the ESO tutorials channel contained 94 videos at the time of writing (June 2026), covering all stages of working with ESO telescopes and services—from observation preparation to accessing the ESO Science Archive and data reduction. Additional content is expected to be added.

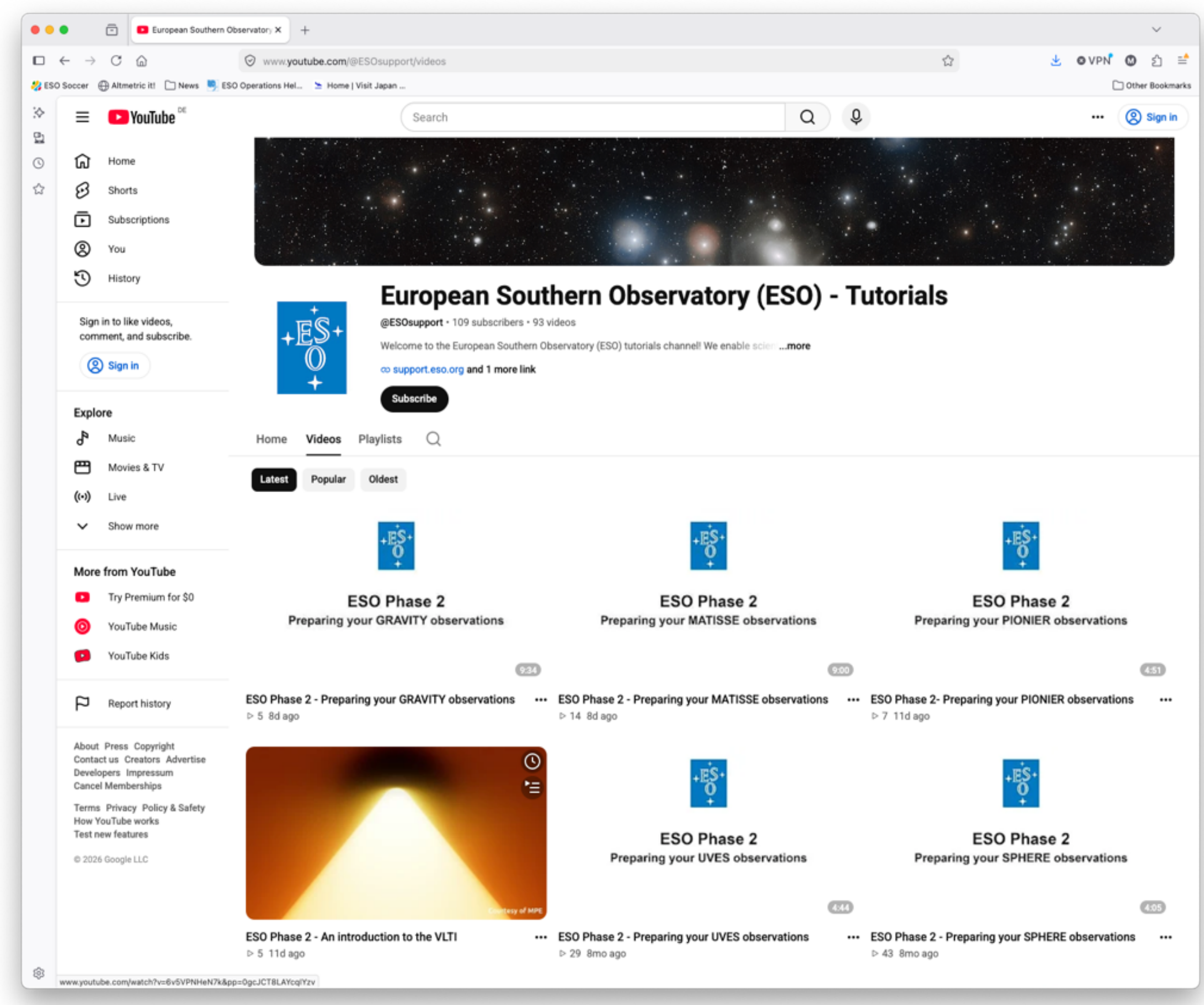

Fig. 11: Screenshot of the ESO tutorials channel (https://www.youtube.com/@ESOsupport) with tutorial videos on VLTI and its PIONIER, MATISSE, and GRAVITY instruments.

[§§§§] https://www.youtube.com/watch?v=6v5VPNHeN7k
[*****] https://www.youtube.com/watch?v=rJi3ZFpNTj0
[†††††] https://www.youtube.com/watch?v=joCOlh8Ac2U
[‡‡‡‡‡] https://www.youtube.com/watch?v=tiqqSTwsOv0

# 8. ARCHIVE DATA PRODUCTS

A further element of the end-to-end support model is the provision of reduced data products. In collaboration with the VLTI Expertise Centers[§§§§§] and the Jean-Marie Mariotti Center (JMMC)[******], ESO is undertaking an ongoing effort to provide GRAVITY interferometric data products through the ESO Science Archive[††††††], with the goal of making VLTI data more accessible and readily usable by the broader community. A large volume of data is already available, including uncalibrated visibilities from the start of operations in 2016 through the current release, which covers data up to 2025-12-31. Figure 12 shows the available setups.

| Field & Axis | Polarisation | Spectral Resolution |
|---|---|---|
| Single-field on-axis | Combine, split | Low ~22, Medium ~500, High ~4000 |
| Dual-field on-axis | Combine, split | Low ~22, Medium ~500, High ~4000 |
| Dual-field off-axis | Combine, split | Low ~22, Medium ~500, High ~4000 |

Fig. 12: Available setups of reduced GRAVITY data in the ESO archive

A forthcoming release of calibrated visibilities will further lower the barrier to scientific exploitation. The data products are generated through homogeneous, automated pipeline processing, ensuring consistency across the dataset. In addition, an automated quality control system provides objective metrics and data quality plots (Figures 13 and 14), enabling users to rapidly assess data quality without the need for full reprocessing. At the time of writing (June 2026), the release[‡‡‡‡‡‡] includes 17,509 uncalibrated visibilities.

Importantly, the GRAVITY data products in the ESO Science Archive are complemented by GRAVITY Jupyter notebooks[§§§§§§], which allow users to explore the data and perform basic analysis steps, including data selection and access, association of science targets with suitable calibrators, pipeline-based calibration, and export to external software such as OIFITS Explorer and LITpro provided by JMMC. Together, these developments enable fast and user-friendly access to reduced interferometric data, significantly lowering the entry barrier for the broader community and increasing the scientific return of VLTI observations.

| Header keyword starts with HIERARCH ESO QC | Description | Bad quality (bright red) | Warning flag (light red) | Good quality |
|---|---|---|---|---|
| FT TRACKING RATIO (in %) | Fringe tracker tracking ratio | < 80 | [80,90] | >90 |
| FT PHASE RMS | Fringe tracker root mean square of the phase | >0.7 | [0.5,0.7] | <0.5 |
| VFACTOR | Ratio of frames non affected from the visibility loss due to flux flickering | <0.45 and >1.1 | [0.45,0.8] | >0.8 |
| PFACTOR | Ratio of frames not affected by the visibility loss due to phase jittering | <0.6 and >1.1 | [0.6,0.85] | >0.85 |
| FT SNRB AVG | SNR for the fringe tracker channel | <10 | [10,30] | >30 |
| SC SNRB AVG | SNR for the science channel | <50 | [50,150] | >150 |

Fig. 13: Quality control parameters of GRAVITY data products in the ESO Science Archive.

§§§§§ https://european-interferometry.eu/vlti-expertise-centers/
****** https://www.jmmc.fr/
†††††† https://archive.eso.org/scienceportal/home
‡‡‡‡‡‡ https://archive.eso.org/scienceportal/home
https://www.eso.org/rm/api/v1/public/releaseDescriptions/242
§§§§§§ https://github.com/eso/astroquery_examples/blob/main/examples/case_studies/gravity_uncalibrated_example.ipynb

PIPEFILE = r.GRAVI.2026-05-31T05:40:14.034_tpl_0000.fits - NCOMBINE: 3
SCIENCE diameter data: from CAT - NAME: HD 139144 - diam: 0.3059014 - err: 0.004164576
FT diameter data: from CAT - NAME: HD 139144 - diam: 0.3059014 - err: 0.004164576
SCIENCE FRAMES REJECTED RATIO (%) : U4-U3=0 - U4-U2=0 - U4-U1=0 - U3-U2=0 - U3-U1=0 - U2-U1=0
FT FRAMES REJECTED RATIO (%) : U4-U3=0 - U4-U2=0 - U4-U1=0 - U3-U2=0 - U3-U1=0 - U2-U1=0

| | U4-U3 | U4-U2 | U4-U1 | U3-U2 | U3-U1 | U2-U1 |
|---|---|---|---|---|---|---|
| FT TRACKING RATIO | 100 | 100 | 100 | 100 | 100 | 100 |
| FT PHASE RMS | 0.32 | 0.33 | 0.32 | 0.34 | 0.33 | 0.32 |
| VFACTOR | 0.89 | 0.9 | 0.91 | 0.89 | 0.89 | 0.91 |
| PFACTOR | 0.98 | 0.99 | 0.99 | 0.98 | 0.98 | 0.99 |
| FT SNRB AVG | 51.4 | 53.0 | 52.9 | 53.2 | 53.1 | 61.6 |
| SC SNRB AVG | 1456.3 | 1393.8 | 1456.3 | 1387.4 | 1609.9 | 1387.4 |
| PBL start to end (m) | 60.7 to 60.6 | 81.7 to 81.3 | 41.3 to 41.0 | 115.1 to 114.4 | 91.4 to 91.0 | 51.0 to 50.8 |
| PBLA start to end (deg) | 129.4 to 130.1 | 100.2 to 100.8 | 54.3 to 54.6 | 77.3 to 77.8 | 45.7 to 46.0 | 38.7 to 39.0 |

PROCSOFT = GRAVITY_ 1.9.7 - PRO CATG = SINGLE_CAL_VIS - PIPEFILE = r.GRAVI.2026-05-31T05:40:14.034_tpl_0000.fits
SOBJ NAME = HD_139144 - SOBJ (K) MAG = 5.9
ROBJ NAME = HD_139144 - ROBJ (K) MAG = 5.9
INSMODE = SINGLE,HIGH,SPLIT,SPLIT - FEED MODE = SINGLE_STS - ROOF POS = ONAXIS
TAU0 (ms) START = 3.9 END = 4.1
FWHM (") START = 0.71 END = 0.76
VIS2ERR = 0.0117 - T3PHIERR (deg) = 0.5568

UT4-U4
FT P05P95 = 0.687
UT3-U3
FT P05P95 = 0.449
UT2-U2
FT P05P95 = 0.693
UT1-U1
FT P05P95 = 0.693
FT FLUXDATA
SC FLUXDATA
VIS2DATA
VISPHI
T3PHI
created: 2026-06-19 15:20:49

Fig. 14: Example data-quality sheet of GRAVITY data products in the ESO Science Archive. For a description of the plot, please see the release description referenced in the main text.

## 9. COMMUNITY CONTRIBUTIONS TO VLTI OPERATIONS

This support model is complemented by an ecosystem of community-developed tools and expertise. VLTI operations have benefited from such community contributions since the start of operations. The JMMC has collaborated closely with ESO on the development of observation preparation tools and calibrator catalogues. JMMC provides a comprehensive suite of tools[*******] [49], covering proposal preparation (e.g. ASPRO2 [50], A2P2, GetStar, SearchCal [51], SearchFFT), data analysis (e.g. OIFITS validator, OIFITS Explorer, LITpro [52], OImaging, AMHRA, WISARD), and databases (e.g. JMDC, JSDC, OIDB, BadCal) for VLTI and other interferometers.
The VLTI Expertise Centers of the European Interferometry Initiative, distributed throughout Europe, provide in-depth support for observation preparation, advanced data reduction, and interpretation [53]. The European Interferometry Initiative also organises VLTI Interferometry Schools and VLTI Community Days, and manages the Fizeau exchange visitor programme in optical interferometry, among other activities.

The ESO User Support Department has developed an LLM prompt for automatic compilation of ESO- and community-provided tools for proposal preparation, data analysis, modelling, and databases across VLT/VLTI instruments and facilities. The resulting inventory, which is curated manually, includes information on each tool, such as type, origin, description, references, links, and status. For VLTI specifically, this approach identified 52 entries, one of the largest numbers across VLT/VLTI facilities, highlighting the strong and ongoing community engagement in interferometry software development and support.

******* https://www.jmmc.fr/english/tools/

After excluding ESO-developed tools such as the data reduction pipelines[†††††††] and inactive software, in particular those associated with decommissioned instruments (e.g. AMBER and MIDI), 26 entries remained: 12 from JMMC, 2 from instrument consortia, and 12 from the broader community. Most of these tools operate at the general interferometry level (20), while others target specific instruments, including GRAVITY (e.g. orbitize! [54], whereistheplanet [55], exogravity [56], GR fit codes [57,58]), MATISSE (e.g. IRBis [59], mat_tools), and PIONIER (pndrs [60]). Additional tools include modelling packages (PMOIRED [61], CANDID [62], oimodeler [63]) and image-reconstruction methods (MiRA [64], BSMEM [65,66], SQUEEZE [67], SPARCO [68], OITOOLS/ROTIR[‡‡‡‡‡‡‡]). Figure 15 shows an example of the resulting inventory of community contributed software.

Overall, there is strong community engagement in providing complementary tools and services for VLTI operations and data analysis. Further integration of such tools into platforms such as ObsPrep and archive interfaces—for example via API-based services or advanced Jupyter notebooks—would enhance usability and accessibility for the community.

| Software | Instrument(s) | Observation type(s) | Origin | Description | Reference | Link | Status | Verification | Notes / overlap |
|---|---|---|---|---|---|---|---|---|---|
| Aspro2 | All VLTI | Preparation | JMMC | Observation preparation: observability, (u,v) coverage, OIFITS simulation, OB generation; SAMP-interoperable; supports GRAVITY/MATISSE differential modes and GPAO. | Bourgès et al. 2014, ASPC | https://www.jmmc.fr/aspro2 | Active (rel. through Dec 2025) | Verified | Writes VLTI/CHARA observing blocks. |
| A2P2 | All VLTI | Preparation | JMMC | Bridge pushing Aspro2 observing blocks to ESO P2 (and CHARA backends). | JMMC | https://www.jmmc.fr | Active | Verified | Overlaps ESO P2. |
| SearchCal | All VLTI | Calibrator | JMMC | Calibrator-search engine building dynamic catalogues of suitable calibrators (bright & faint cases) with predicted diameters/visibilities. | Bonneau et al. 2006, A&A 456, 789; 2011, A&A 535, A53 | https://www.jmmc.fr | Active | Verified | Feeds Aspro2 via SAMP. |
| GetStar | All VLTI | Calibrator | JMMC | Single-target service returning computed stellar parameters/diameters (AllWISE/Gaia); now Aspro2's default resolver. | JMMC | https://www.jmmc.fr | Active | Verified | Generalises SearchCal. |
| JSDC | All VLTI | Calibrator | JMMC | JMMC Stellar Diameters Catalog queried by SearchCal/GetStar. | Lafrasse et al. 2010; Chelli et al. 2016, A&A 589, A112 | | Active | Verified | VizieR II/346. |
| BadCal | All VLTI | Calibrator | JMMC | Community list of known-bad (resolved/binary) calibrators to avoid. | JMMC | https://www.jmmc.fr | Active | Verified | |
| OiDB | All VLTI | Archive | JMMC | Optical Interferometry Database: archive/search of reduced OIFITS; sole public home of pre-2014 visitor-mode PIONIER/AMBER products. | JMMC | https://oidb.jmmc.fr | Active | Verified | Complements the ESO archive. |
| AMHRA | All VLTI | Models | JMMC | Serves astrophysical model image-cubes (disks, fast rotators...) as OIFITS/FITS into Aspro2/OImaging. | JMMC | https://www.jmmc.fr | Active | Verified | |
| PMOIRED | All VLTI | Spectro-interferometry | Community | Python library to visualise/fit OIFITS with composable geometric + chromatic components; handles correlated errors; bootstrap & grid fitting. | Mérand 2022, ascl:2205.001 | https://github.com/amerand/PMOIRED | Very active (PyPI v1.3.16, Jun 2026) | Verified | De-facto Python standard for parametric VLTI fitting. |
| CANDID | All VLTI | Binary detection | Community | Systematic χ² grid search for faint companions + detection-limit maps from OIFITS. | Gallenne et al. 2015, A&A 579, A68 | https://github.com/amerand/CANDID | Active | Verified | By Gallenne, Mérand, Kervella. |
| oimodeler | All VLTI | Spectro-interferometry | Community | Modular Python modelling: analytic Fourier + image-plane + radiative-transfer-cube components, chromatic & time-dependent, emcee fitting. | Meilland et al. 2024, SPIE 13095 | https://gitlab.oca.eu/oimodeler/oimodeler | Active (GitLab v1.0; PyPI lags 0.8.1) | Verified | Co-developed in MATISSE modelling group. |
| LITpro | All VLTI | Model fitting | JMMC | GUI/engine for least-squares model-fitting of OIFITS from geometric building blocks; SAMP-integrated with Aspro2. | Tallon-Bosc et al. 2008, SPIE 7013 | https://www.jmmc.fr | Active | Verified | |
| MiRA | All VLTI | Imaging | Community | Multi-aperture Image Reconstruction Algorithm (Yorick); regularised-likelihood gradient minimisation; interfaced as a recipe inside the GRAVITY pipeline. | Thiébaut 2008, SPIE 7013 | https://github.com/emmt/MiRA | Active | Verified | Also reachable via OImaging. |
| WISARD | All VLTI | Imaging | JMMC | Self-calibration / weak-phase reconstruction (IDL/GDL). | Meimon, Mugnier & Le Besnerais 2005 | https://www.jmmc.fr/wisard_page.htm | Maintained (JMMC) | Verified | ONERA → JMMC. |
| BSMEM | All VLTI | Imaging | Community | BiSpectrum Maximum Entropy Method reconstruction. | Buscher 1994; Baron & Young 2008 | | Available | Verified | Widely used in imaging contests. |
| SQUEEZE | All VLTI | Imaging | Community | MCMC / parallel-tempering reconstruction with error bars; polychromatic; optional simultaneous analytic (SPARCO) modelling. | Baron, Monnier & Kloppenborg 2010, SPIE 7734 | https://www.chara.gsu.edu/analysis-software/imaging-software | Active (GPLv3) | Verified | F. Baron, GSU. |
| IRBis | MATISSE; All VLTI | Imaging | Community | Image reconstruction (ASA_CG optimiser); shipped as part of the MATISSE DRS and exposed via mat_tools. | Hofmann, Weigelt & Schertl 2014, A&A 565, A48 | | Active (within MATISSE) | Verified | Hofmann & Weigelt, MPIfR. |
| SPARCO | All VLTI | Imaging | Community | Semi-parametric chromatic reconstruction separating point-source from extended emission; used as a mode in MiRA/SQUEEZE. | Kluska et al. 2014, A&A 564, A80 | | Active (method/add-on) | Verified | |
| OImaging | All VLTI | Imaging | JMMC | Unified GUI front-end driving MiRA/BSMEM/WISARD/SPARCO on OIFITS. | JMMC | https://www.jmmc.fr | Active | Verified | Wrapper over reconstruction engines. |
| OITOOLS / ROTIR | All VLTI | Imaging / model | Community | Julia environment to load/simulate/model-fit/image OI data (CHARA/NPOI/VLTI); ROTIR images rotating-star surfaces. | F. Baron (Julia) | https://www.chara.gsu.edu/analysis-software/imaging-software | Active | Verified | |
| orbitize! | GRAVITY (+ direct imaging) | Astrometry / RV | Community | Bayesian orbit-fitting for relative-astrometry companions; standard for combining GRAVITY astrometry with imaging/RV. | Blunt et al. 2020, AJ 159, 89 | https://orbitize.info | Very active (PyPI v3.3.2, Jun 2026) | Verified | |
| whereistheplanet | GRAVITY (+ direct imaging) | Astrometry | Community | Predicts companion sky positions from fitted orbits — used to plan GRAVITY dual-field pointings. | Wang et al. 2021 | | Active | Verified | GitHub + web app. |
| exogravity | GRAVITY | High-contrast / exoplanet | Community | Open-source pipeline isolating a companion's flux from residual starlight in dual-field data; builds the ExoGRAVITY K-band Spectral Library. | Nowak et al. 2024; GRAVITY Collab. 2020, A&A 633, A110 | | Active | Verified | Operates on pipeline 'astroreduced' products. |
| GRAVITY Galactic-Centre / GR-fit codes | GRAVITY | Galactic Centre / astrometry | Consortium | S2-orbit, Schwarzschild-precession and Sgr A* flare-astrometry fitting behind the Galactic-Centre results. | GRAVITY Collab. 2018/2020 (A&A 615, L15; 636, L5) | | Paper-internal | Needs verification | Bespoke MCMC; no released package located. |
| mat_tools | MATISSE | All | Consortium | Python wrapper/visualisation around the ESO DRS: mat_autoCalib, mat_fluxCal, mat_mergeAllOiFits (BCD merge), plus the IRBis interface. | Consortium | https://github.com/Matisse-Consortium/tools | Active | Verified | Consortium-supported. |
| pndrs | PIONIER | Interferometry, imaging | Community / JMMC | PIONIER Data Reduction Software (Yorick): raw scans → calibrated V² & closure phases in OIFITS; FLUOR-derived estimator. | Le Bouquin et al. 2011, A&A 535, A67 | https://www.jmmc.fr/english/tools/data-processing/pionier-68/ | Active (≥ v3.74) | Verified | JMMC account required to download. |

Fig. 15: Example inventory of community-contributed tools related to VLTI and optical/infrared interferometry.

††††††† https://www.eso.org/sci/software/pipe_aem_main.html
‡‡‡‡‡‡‡ https://chara.gsu.edu/analysis-software/imaging-software

## 10. OPERATIONS FOR THE FUTURE OF INTERFEROMETRY

Based on the experience with VLTI operations and the operational developments described above, several areas for improvement can be identified: some applicable on relatively short time scales for VLTI operations, while others should be considered on longer time scales for new facilities.

### Short-term developments and improvements

Further development of observation preparation tools remains a priority. In particular, extending the capabilities of ObsPrep will allow users to define observations at a higher level, while the tool selects instrument configurations and prepares the corresponding observing blocks. This will reduce the complexity of Phase 2 preparation and improve the quality of the prepared material. This aligns with the general roadmap for Phase 2 preparation and ObsPrep for the VLT and toward the ELT.

In addition, the continued integration of functionalities into the Phase 2 tools is expected to improve efficiency. This includes enhanced finding-chart generation, improved support for the selection of calibrators and guide stars, and the provision of *uv*-coverage visualisation for PIs during an observing period and for archive users after the observations have been completed.

Another area of short-term development is the continued expansion of archived data products. The availability of reduced interferometric data already lowers the barrier for scientific exploitation. Further improvements in standardised pipeline processing for all VLTI instruments, including data calibration, automated quality control, and user-facing analysis tools such as Jupyter notebooks, will increase the usability of these data products.

Finally, a large number of community-developed tools exist for proposal preparation, data analysis, and modelling. At present, these tools are distributed across different platforms. Their integration into common environments, such as ObsPrep or archive interfaces, would simplify access and provide a more coherent workflow for users.

### Lessons learned and recommendations for future interferometric facilities

The experience gained from VLTI operations highlights several points that are relevant for future interferometric facilities. A central aspect is the importance of embedding operations within a consistent end-to-end data flow model. VLTI has benefited enormously from being integrated into the overall VLT end-to-end system, in which proposal preparation, scheduling, observation execution, data reduction, quality control, and archive access are closely connected. This integration has been essential for efficient operations and for ensuring consistency of data products. Similar end-to-end concepts should therefore be considered as a baseline requirement for future facilities.

The use of movable telescopes has enabled a wide range of baseline configurations and has been essential for interferometric imaging and for sampling the *uv* plane. At the same time, this flexibility introduces operational constraints. Telescope relocations require additional overheads and lead to a more fragmented and less flexible observing schedule compared to facilities with fixed configurations. Future facilities should carefully consider this trade-off and ensure that scheduling tools and operational concepts are adapted accordingly.

A second important aspect is the coexistence of different observation types, such as imaging, astrometry, time-series, and snapshot observations. These observing modes have different and sometimes competing requirements in terms of scheduling, baseline configurations, and operational support. For example, imaging benefits from rapid changes of configurations and dense *uv* coverage, while time-series observations require regular repetition of specific configurations. Astrometry often also requires time-series capabilities, baselines suitable for high-precision astrometry, and dedicated support for advanced observation preparation and data analysis. Future facilities should therefore clearly define whether operations are optimised for a specific observing mode, or whether multiple observation types are to be supported in parallel. In the latter case, the operations concept should explicitly address how these potentially conflicting requirements are prioritised and accommodated.

A further important lesson is the strong contribution of the community to VLTI operations. External tools, expertise centres, and coordinated initiatives have played a significant role in observation preparation, data analysis, and user support. These contributions have been essential for the scientific productivity of the facility. A more systematic integration of community-developed tools into shared platforms would further improve accessibility and reduce duplication of effort.

Looking further ahead, the increasing complexity of interferometric data will also connect future operations to broader developments in scientific computing infrastructure. This may raise questions about the suitability of existing data formats and analysis workflows. While current formats such as OIFITS have proven effective, future developments, including cloud-based data processing and large-scale analysis, may require adaptations. This includes considerations of data structure, metadata completeness, and interoperability with modern computing environments. These aspects should be taken into account in the design of future facilities and data systems.

Overall, the recent evolution of VLTI operations shows that increasing interferometric capability must be accompanied by an equally strong evolution of user support. Higher sensitivity, more complex observing modes, and broader scientific use require support across the full observing cycle, from proposal and Phase 2 preparation to scheduling, execution, archive data products, and community tools. The VLTI experience demonstrates that such an integrated operational model is essential for maintaining broad community access to complex interferometric facilities. These lessons should provide useful guidance for the design and operation of future optical/infrared interferometers.

## ACKNOWLEDGEMENTS

We acknowledge the community interest and engagement in VLTI and contributions from instrumentation teams and centres of expertise, which led not only to enjoyable and fruitful collaborations, but also resulted in the state-of-the-art facilities and operations that serve ever growing VLTI scientific community.